# Superconducting Two-Dimensional Metal-Organic Framework

Xiaoming Zhang[1,2], Yinong Zhou[2], Bin Cui[1,2], Mingwen Zhao[1*] and Feng Liu[2,3*]

[1]School of Physics and State Key Laboratory of Crystal Materials, Shandong University, Jinan, Shandong 250100, China.

[2]Department of Materials Science and Engineering, University of Utah, Salt Lake City, Utah 84112, USA

[3]Collaborative Innovation Center of Quantum Matter, Beijing 100084, China

*Correspondence to: fliu@eng.utah.edu; zmw@sdu.edu.cn

ABSTRACT

Superconductivity is a fascinating quantum phenomenon characterized by zero electrical resistance and the Meissner effect. To date, several distinct families of superconductors (SCs) have been discovered. These include three-dimensional (3D) bulk SCs in both inorganic and organic materials as well as two-dimensional (2D) thin film SCs but only in *inorganic* materials. Here we predict superconductivity in 2D and 3D *organic* metal-organic frameworks by using first-principles calculations. We show that the highly conductive and recently synthesized Cu-benzenehexathial (BHT) is a Bardeen-Cooper-Schrieffer SC. Remarkably, the monolayer Cu-BHT has a critical temperature ($T_c$) of 4.43 K while $T_c$ of bulk Cu-BHT is 1.58 K. Different from the enhanced $T_c$ in 2D inorganic SCs which is induced by interfacial effects, the $T_c$ enhancement in this 2D organic SC is revealed to be the out-of-plane soft-mode vibrations, analogous to surface mode enhancement originally proposed by Ginzburg. Our findings not only shed new light on better understanding 2D superconductivity, but also open a new direction to search for SCs by interface engineering with organic materials.

Since the first discovery of superconductor (SC) a century ago [1], the field of superconductivity research has been continuously filled with exciting discoveries of new SCs. The first generation of SCs includes metal conductors, either elemental metal [1], metal alloys [2], or metallic phase of pressurized semiconductors [3]. They are conventional Bardeen-Cooper-Schrieffer (BCS) [4] SCs with a relatively low critical temperature ($T_c$) determined by electron-phonon interaction. The second generation came as a breakthrough in the field with the discovery of high-$T_c$ coper oxides SCs [5,6]. The third-generation SCs is a family of Fe-based pnictide compounds [7].

For all the three families of 3D bulk SCs, their 2D thin-film counterparts have also been actively pursued [8-10], especially with the recent advance in epitaxial growth of thin films with single atomic layer precision. 2D SCs are appealing for fundamental study, with interesting properties pertaining to 2D geometry, and have potential applications for making 2D SC devices. It is expected that the superconductivity may interplay with other quantum effects at the 2D limit, such as charge localization [11], quantum size effect [12], and topological and quantum phase transitions [13-15]. Recent studies have found some 2D SCs, such as $La_{2-x} Sr_xCuO/La_2CuO_{4+\delta}$ interface [16] [21] and FeSe single layer [17], exhibit a $T_c$ much higher than their 3D counterparts owing to interfacial phonon mode coupling [18] and electrical field effect [19]. However, all the known thin-film and 2D SCs are *inorganic* materials [8-10].

Here, we predict for the first time superconductivity in both 2D and 3D *organic* metal-organic frameworks (MOFs). Using first-principles calculations, we show that single-atomic-layer Cu-benzenehexathial (Cu-BHT) is a BCS 2D SC with a $T_c$ of 4.43 K, while the 3D Cu-BHT has a $T_c$ of 1.58 K. We reveal that the enhanced $T_c$ in the single layer Cu-BHT is caused by the out-of-plane soft-mode vibrations allowed at the 2D limit, consistent with the softening of surface modes to enhance $T_c$ of thin-film SCs as envisioned by Ginzburg [20]. We emphasize that it is mechanistically different from interfacial effects as discovered in 2D inorganic SCs [18,19].

Our study is largely motivated by the recent discovery of surprisingly high room-temperature electrical conductivity up to 1580 s · cm$^{-1}$ in Cu-BHT [21], in contrast to conventional MOFs which are commonly insulating or semiconducting. In principle, any nonmagnetic metal could become superconducting at sufficiently low temperature. Therefore, this experimental observation intrigued us to investigate the possible superconductivity in Cu-BHT. Furthermore, because Cu-BHT has a layered structure, we want to find out whether this unique material, with the highest conductivity among all the known MOFs to date, [21] will offer the first 2D organic SC.

The atomic structure of single layer Cu-BHT is shown in Fig. 1a. Copper atoms are bridged by BHTs via Cu-S bonds, forming a Kagome lattice. The calculated equilibrium lattice constant $a_0$ is 8.80 Å, in good agreement with the previous calculation (8.76 Å) [21]. From the orbital-resolved band structures of Cu-BHT shown in Fig. 1b (also see Supplemental Material, Fig. S1a and Fig. S1b [22]), one can clearly see the intrinsic metallic nature with multiple bands crossing the Fermi level. These bands are characterized with symmetry-guided coordination between $S_{p_x+p_y}$ orbitals (Fig. 1b) and $Cu_{d_{xy}+d_{x^2-y^2}}$ orbitals (see Supplemental Material, Fig. S1b [22]), and between $S_{p_z}$ orbitals (Fig. 1b), $C_{p_z}$ orbitals (see Supplemental Material, Fig. S1a [22]) and $Cu_{d_{xz}+d_{yz}}$ orbitals (see Supplemental Material, Fig. S1b [22]). Together they form a high density of 2D $\pi$-$d$ electron gas at the Fermi level.

The Fermi surface of single layer Cu-BHT is shown in Fig. 2a, where one can see several hole/electron pockets in the Brillouin zone (BZ). The two hole pockets centered at Γ point have the characteristics of **π**-bands stemming from the $p_z$ orbitals of C and S atoms along with the $d_{xz}+d_{yz}$ orbitals of Cu, which can be seen more clearly from the partial charge density shown in Fig. S1c of Supplemental Material [22]. An electron pocket at the K point and a hole pocket at the M point around the boundary of BZ have the **σ**-band characteristics arising from $p_x+p_y$ orbitals of S and $d_{xy}+d_{x^2-y^2}$ orbitals of Cu atoms, as shown in Fig. S1d of Supplemental Material [22]. Figure 2b shows the phonon dispersion $\omega_{\mathbf{q}\nu}$ with the frequency lower than 450 cm$^{-1}$, which confirms the dynamic stability of 2D Cu-BHT by the absence of imaginary frequency modes. The phonon dispersion over whole frequency range and the projected phonon density of state (PhDOS) are displayed in Fig. S2 of Supplemental Material [22]. Both Cu and S atoms dominate the vibration modes with frequencies below 450 cm$^{-1}$. The interaction between C and S atoms contributes to the intermediate-frequency region between 450 and 1000 cm$^{-1}$. The in-plane stretching modes of C atoms occupy the high frequencies above 1000 cm$^{-1}$.

Figure 2b also shows the calculated electron-phonon coupling (EPC) strength $\lambda_{\mathbf{q}\nu}$ (red circles),

$$\lambda_{\mathbf{q}\nu} = \frac{4}{\omega_{\mathbf{q}\nu} N(0) N_{\mathbf{k}}} \sum_{\mathbf{k},n,m} \left| g^{\nu}_{\mathbf{k}n,\mathbf{k}+\mathbf{q}m} \right|^2 \delta(\varepsilon_{\mathbf{k}n}) \delta(\varepsilon_{\mathbf{k}+\mathbf{q}m})$$

$$g^{\nu}_{\mathbf{k}n,\mathbf{k}+\mathbf{q}m} = \frac{\langle \mathbf{k}n | \delta V / \delta u_{\mathbf{q}\nu} | \mathbf{k}+\mathbf{q}m \rangle}{\sqrt{2\omega_{\mathbf{q}\nu}}},$$

where $N(0)$ and $N_{\mathbf{k}}$ are respectively the electron DOS and the number of $\mathbf{k}$ points at the Fermi level. $u_{\mathbf{q}\nu}$ is amplitude of the displacement of phonon and $V$ is the Kohn-Sham potential. $g^{\nu}_{\mathbf{k}n,\mathbf{k}+\mathbf{q}m}$ is the scattering amplitude of an electronic state $|\mathbf{k}+\mathbf{q}m\rangle$ into another state $|\mathbf{k}n\rangle$ resulting from the change in the self-consistent field potential $\delta V/\delta u_{\mathbf{q}\nu}$ arising from a phonon $\omega_{\mathbf{q}\nu}$. The size of the red circles in Fig. 2b is drawn proportional to the EPC $\lambda_{\mathbf{q}\nu}$ for the corresponding phonon mode. One can clearly see that the low-energy optical branches with the frequencies around 50 cm$^{-1}$ possess the strongest electron-phonon couplings, while contribution from other branches is negligible. Moreover, the softening of one acoustic branch yields significant coupling between electrons and acoustic phonons. Knowing the main phonon modes that couple with electron/hole, one can reveal the distribution of electron-phonon couplings over the whole BZ and hence determine the contribution of the electron/hole pockets at the Fermi level. The momentum-dependent coupling $\lambda_{\mathbf{q}}$, as calculated by $\lambda_{\mathbf{q}} = \sum_{\nu} \lambda_{\mathbf{q}\nu}$, is plotted in Fig. 2c. It reveals that the electron/hole pockets with σ-band characteristics at the boundary of BZ contribute predominantly to the EPC, while the hole pockets with the π-band characteristics centered at Γ point make relatively small contributions.

Next, we compare the Eliashberg function $\alpha^2 F(\omega)$, the cumulative frequency-dependent EPC function $\lambda(\omega)$, and the displacement decomposed partial PhDOS to analyze the contributions of directional atomic vibration to the EPC constant.

$$\alpha^2 F(\omega) = \frac{1}{2N_\mathbf{q}} \sum_{\mathbf{q}\nu} \lambda_{\mathbf{q}\nu} \omega_{\mathbf{q}\nu} \delta(\omega - \omega_{\mathbf{q}\nu})$$

$$\lambda(\omega) = 2 \int_0^\omega \frac{\alpha^2 F(\omega')}{\omega'} d\omega' \ .$$

Figure 3 clearly indicates that the predominant contribution is from the out-of-plane vibrations of S atoms, which lead to a sharp peak in $\alpha^2 F(\omega)$ at the frequencies around 50 cm$^{-1}$. Another contribution stems from the out-of-plane vibrations of Cu atoms, which result in a small increment of EPC at the narrow frequency range, around 100 cm$^{-1}$. Given the electronic states of S and Cu atoms at the Fermi level, the out-of-plane vibrations couple with the σ-band characteristic electron/hole pockets at the Fermi level, which is consistent with the distributions of $\lambda_\mathbf{q}$ shown in Fig. 2c. Even through symmetry allows the coupling between the π-band characteristic hole pockets and in-plane optical stretching modes of carbon atoms, the high frequency (larger than 1000 cm$^{-1}$) of such stretching mode makes it almost negligible, similar to graphene [23]. The total EPC is found to be λ = 1.16 when $\omega \to \infty$, which makes the 2D single layer Cu-BHT an intermediate to strong conventional superconductor. This large value of λ is likely resulted from a very high density of states with multiple electron/hole packets forming time reversal pairs at the Fermi level (see Fig. 2a) to facilitate a strong electron-phonon coupling (see Fig. 2c).

Using the above values, the logarithmically averaged frequency defined as $\langle\omega\rangle_{\log} = \exp\left[\frac{2}{\lambda} \int \frac{d\omega}{\omega} \alpha^2 F(\omega) \log \omega\right]$ is calculated to be 51.8 K or 35.96 cm$^{-1}$. The T$_c$ is estimated according to the Allen-Dynes [24] modified McMillan's [25] formula:

$$T_c = \frac{\langle\omega\rangle_{\log}}{1.2} \exp\left[\frac{-1.04 \times (1+\lambda)}{\lambda - \mu^* \times (1+0.62\lambda)}\right] \ ,$$

here $\mu^*$ is the retarded Coulomb potential. Using the typical value of $\mu^* = 0.1$, the above formula gives $T_c = 4.43$ K, which is comparable to that of single atomic layer of inorganic metal film [12].

It would be interesting to know whether bulk Cu-BHT is also a SC, and if so, how its T$_c$ compared with a single layer Cu-BHT? The crystal structure of bulk Cu-BHT is presented in Fig. S3 of Supplemental Material [22], which shows a slight slippage between the neighboring two layers. One significant impact caused by the bulk interlayer interaction is hardening of the low-energy optical and acoustic branches, which suppress the coupling between σ-band states and out-of-plane vibrations, as shown in Fig. 4a. On the other hand, the interlayer interaction enhances the EPC along the Γ-Z direction, similar to that in GIC [26,27]. However, overall EPC is significantly suppressed, leading to a drop of the first dominant peak in Eliashberg function $\alpha^2 F(\omega)$, as shown in Fig. 4b. The calculated EPC constant and the logarithmically averaged frequency for bulk Cu-BHT is 0.51 and 122.7 K, respectively, which give $T_c$ = 1.58 K as estimated by the McMillan-Allen-Dynes formula. In terms of electronic states, going from 2D to bulk Cu-BHT, both the breakage of π-band states in carbon rings and the weakening of S-Cu σ-band states at Fermi level (see Supplemental Material, right panel of Fig. S3 [22]) reduce the effective DOS $N(0)$ that couple with phonons, leading to fewer coupling modes available. This effect is similar to $LiC_6$ where the $T_c$ of Li-intercalated graphite is lower than that of Li-intercalated graphene due to the reduction of electronic states of Li at the Fermi level [28].

We stress that the finding of bulk Cu-BHT being superconducting is a significant result in its own right, as it represents the discovery of a new SC in organic materials of MOFs by theory. In fact, we expect that experimental confirmation of our theoretical prediction will likely come with bulk measurements first, since the metal-like high conductivity has already been observed at high temperature in the existing bulk samples [21], and efforts are being made to prepare high-quality crystals and larger crystalline domains, as well as thin film and 2D samples [21].

Finally, we point out that the study of superconductivity in Cu-BHT and related MOFs not only opens a new avenue towards 2D and 3D *organic* SCs but may also shed new light on better understanding the existing 2D *inorganic* SCs. We notice that similar 2D in-plane coordination of Cu-S, Cu-O and Fe-Se bonding networks exist in Cu-BHT, Cu-O plane in high $T_c$ cuprate [6] and single layer FeSe [17], respectively, and they all play the key role in trigging the superconductivity in each system. Their Fermi surface are similar having electron and/or hole pockets formed by *d*-orbitals of transition metals and *p*-orbitals of group-VI elements at high symmetry K points. $T_c$ is expected to be strongly affected by doping, strain and substrate (interface), which propose interesting subjects for future studies. Furthermore, the discovery of superconductivity in a Kagome lattice is very interesting from the symmetry point view, as the Kagome lattice is the most intriguing playground for many exotic phenomena, such as fractional topological states [29,30] and quantum spin liquid phases [31,32].

Y. Z. and F.L. acknowledge financial support from DOE-BES (No. DE-FG02-04ER46148). M.W.Z. acknowledge financial support from the National Natural Science Foundation of China (No. 21433006) and the National Key Research and Development Program of China (No. 2016YFA0301200). X.M.Z. and B.C. acknowledge support by the China Scholarship Council (No. 201506220046). We also thank Supercomputing Center at NERSC and CHPC at University of Utah for providing the computing resources.

**Figures**

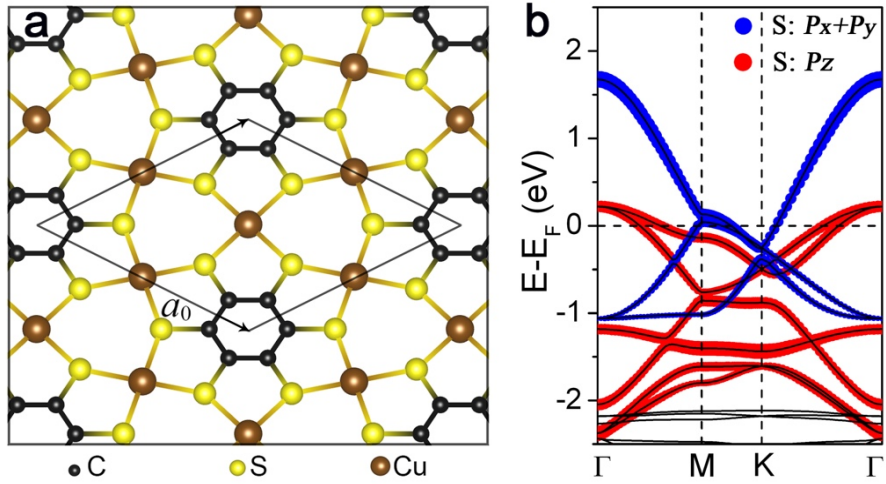

Figure 1 (color online). Crystal and band structure of single layer Cu-BHT. (a) Schematic of crystal structure with the unit cell indicated by a rhombus. $a_0$ is the lattice constant. (b) Band structure with the contributions of $p_z$ and $p_x+p_y$ orbitals of S atoms being indicated by red and blue disks.

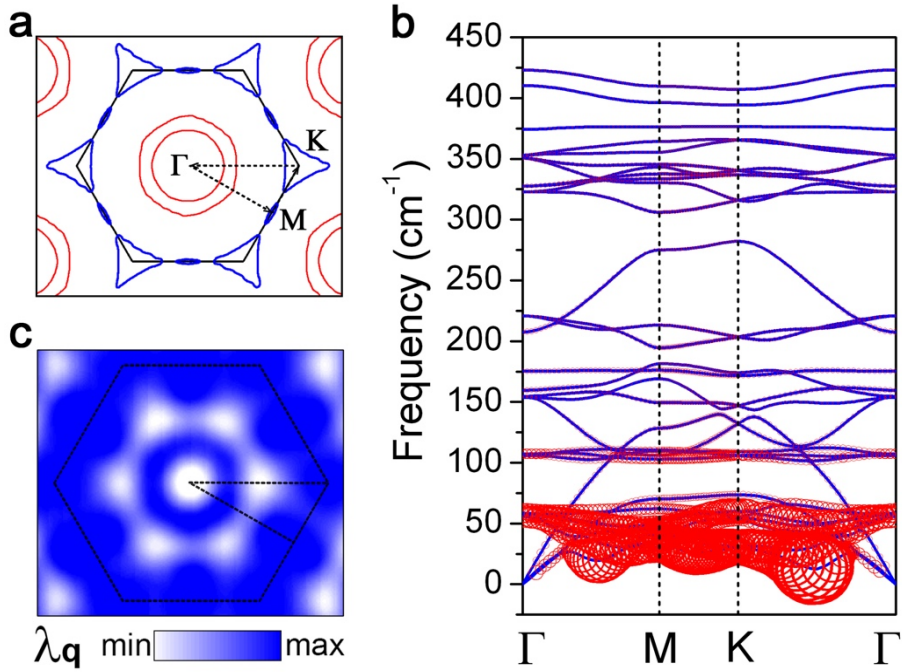

Figure 2 (color online). Fermi surface contour, Phonon spectrum, and Electron-phonon coupling of single layer Cu-BHT. (a) Fermi surface contour in the reciprocal space. (b) Phonon spectrum with the size of red circles being drawn proportional to the magnitude of EPC $\lambda_{\mathbf{q}\nu}$. (c) The distribution of **q**-resolved electron-phonon coupling $\lambda_{\mathbf{q}}$.

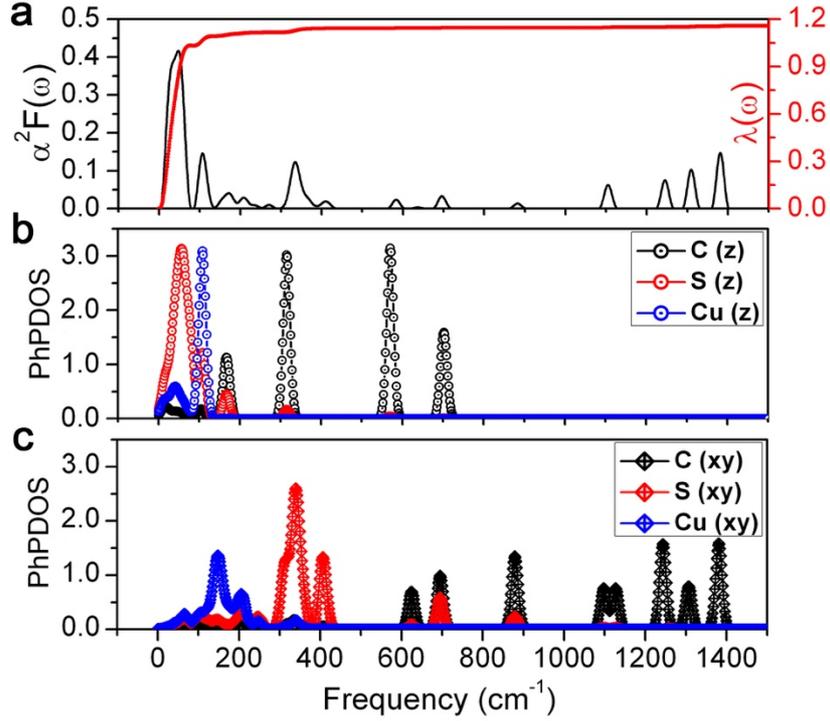

Figure 3 (color online). Eliashberg spectral function and partial phonon density of state (PhDOS) of single layer Cu-BHT. (a) Eliashberg spectral function $\alpha^2F(\omega)$ and cumulative frequency-dependent $\lambda(\omega)$. Partial PhDOS projected on (b) the out-of-plane (z, perpendicular to the Cu-BHT plane) and (c) the in-plane (xy, parallel to the Cu-BHT plane) vibrations of C, S, and Cu atoms.

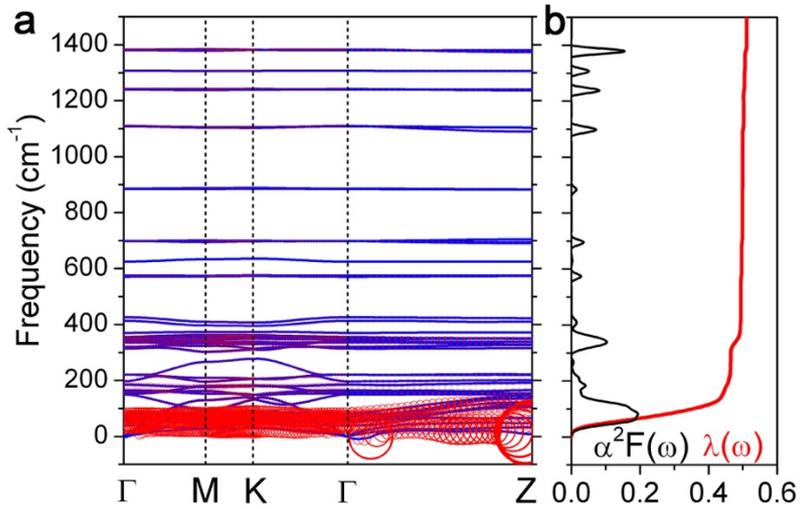

Figure 4 (color online). The superconducting properties of bulk Cu-BHT. (a) Phonon spectrum with the electron-phonon couplings $\lambda_{\mathbf{q}\nu}$ indicated by the red circles. (b) Eliashberg spectral function $\alpha^2F(\omega)$ and cumulative frequency-dependent $\lambda(\omega)$.

*Supplemental Material for*

Superconducting Two-Dimensional Metal-Organic Framework

Xiaoming Zhang[1,2], Yinong Zhou[2], Bin Cui[1,2], Mingwen Zhao[1*] and Feng Liu[2,3*]

[1]School of Physics and State Key Laboratory of Crystal Materials, Shandong University, Jinan, Shandong 250100, China.

[2]Department of Materials Science and Engineering, University of Utah, Salt Lake City, Utah 84112, USA

[3]Collaborative Innovation Center of Quantum Matter, Beijing 100084, China

*Correspondence to: fliu@eng.utah.edu; zmw@sdu.edu.cn

## I. Computational Methods

The calculations are performed using density-functional theory with generalized gradient approximation in the form of Perdew-Burke-Ernzerhof. [1] To eliminate the interlayer interaction, single layer Cu-BHT is simulated by introducing a vacuum layer larger than 15 Å. The superconducting properties were calculated by QUANTUM ESPRESSO code [2] using the norm-conserving pseudopotentials with the cutoff energy of 100 Ry for the wave functions, and the BZ sampling of 8×8×1 (4×4×8 for bulk Cu-BHT) Monkhorst-Pack **k**-point grids with Methfessel-Paxton smearing of 0.07 Ry. Using the above **k**-point samplings for self-consistent cycle, the dynamic matrixes are calculated on a 4×4×1 (4×4×8 for bulk Cu-BHT) **q**-point mesh. The electronic band structures, orbital contributions, and charge density isosurfaces were analyzed by Vienna *ab initio* simulation package [3] with the energy cutoff of 500 eV being used on a 7×7×1 (5×5×9 for bulk Cu-BHT) Monkhorst-Pack sampling in BZ.

## II. Electronic property and Phonon spectrum of single layer Cu-BHT

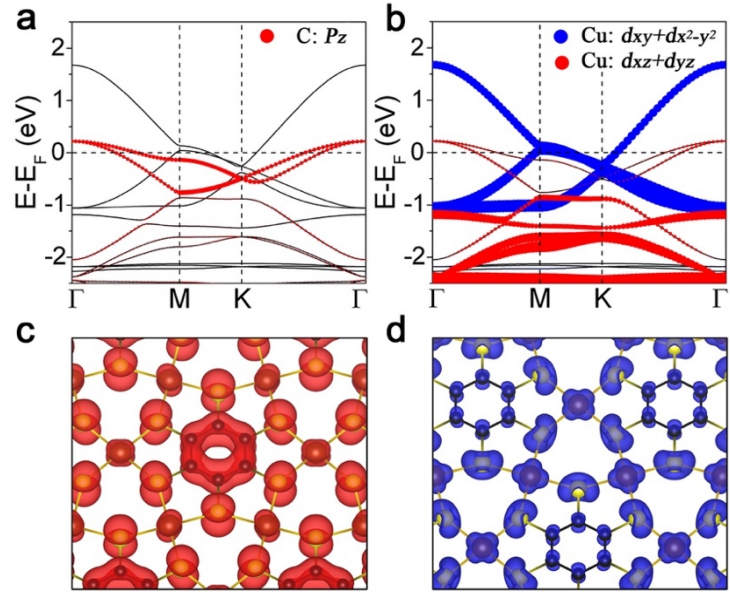

**Figure S1.** Band structures of single layer Cu-BHT with the contributions of the $p_z$ orbitals of C atoms and the *d* orbitals of Cu atoms being indicated in **a**) and **b**), respectively. The partial charge density isosurfaces decomposed to the hole and electron/hole pockets centered at the **c**) center of BZ (Γ point) and **d**) boundary of BZ (K/M point), respectively.

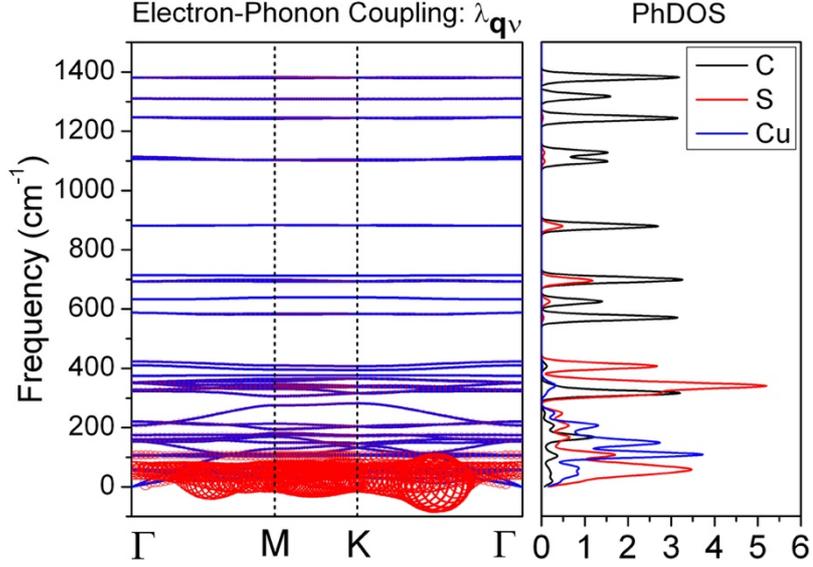

**Figure S2.** Phonon spectrum over whole frequency range and the PhDOS decomposed to the C, S, and Cu atoms of single layer Cu-BHT. The size of red circles, superimposed onto the phonon dispersions with branch index $v$ and wavevector **q**, are drawn in proportion to the magnitude of electron-phonon couplings $\lambda_{\mathbf{q}v}$.

### III. Equilibrium State of bulk Cu-BHT

For the bulk structure of Cu-BHT, the important structural parameters are lattice constant (LC), interlayer distance (ID), and interlayer slip distance (ISD). Because bulk Cu-BHT has a layered structure with weak Van der Waals interlayer interaction, the difference between the LC of single layer Cu-BHT and that of bulk Cu-BHT is expected to be small. Consequently, we used the LC of single layer Cu-BHT in the first-principles calculations to search for the ground-state structure of bulk Cu-BHT as a function of ID and ISD. The structure with the lowest total energy is shown in Fig. S3, which does not have exactly a standard AA or AB stacking. However, the AA and AB stacking patterns reported in Ref. [4]

are both very similar to what we presented here in terms of the relationship between the neighboring two layers. Moreover, the Pawley-refined XRD pattern, the potential energy surface (PES), and the DOS of AA and AB stacking Cu-BHT in Ref. [4] exhibit high similarity, suggesting there is possibly only one stacking pattern for the bulk Cu-BHT as we obtained here.

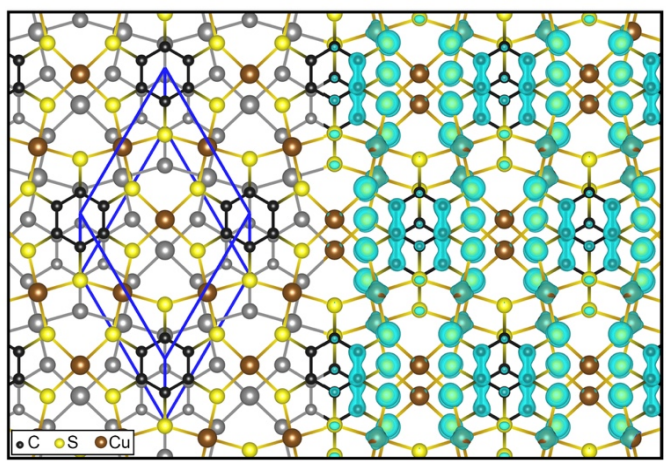

**Figure S3.** The crystal structure of bulk Cu-BHT (left panel) and the partial charge density isosurfaces (right panel) decomposed into the energy range [$E_F$-0.1, $E_F$+0.1] near the Fermi level.

Due to the limitation of computing resource, we adopted the unit cell of AA stacking Cu-BHT as given in Ref. [4] for electron-phonon coupling calculations. The right panel of Fig. S3 shows the partial charge density isosurfaces decomposed to the energy range [$E_F$-0.1, $E_F$+0.1]. Besides the breakage of **π**-band states on carbon rings, the contribution from the $p_{x,y}$ (*d*) orbitals of some S (Cu) atoms disappears, weakening the S-Cu **σ**-band states at the Fermi level.